\documentclass[useAMS,usenatbib]{mn2e}
\usepackage{times,graphicx}
\usepackage{mathrsfs}
\usepackage{epsfig,wrapfig}
\usepackage{mathrsfs}
\usepackage{color}
\usepackage{epsfig,wrapfig}
\usepackage{fancybox}
\usepackage{type1cm}

\newcommand{\degree}{\hbox{$^\circ$$$}}

\renewcommand{\theequation}{\textcolor{blue}{\arabic{equation}}}
\renewcommand{\thefigure}{(\textcolor{red}{\arabic{figure}})}
\renewcommand{\thetable}{(\textcolor{red}{\Roman{table}})}

\title[NASA \textit{\&} ESA Spacecrafts are defying both Newtonian  \textit{\&} Einsteinian Gravitation. Why?]{Are Flyby Anomalies an ASTG Phenomenon? {(ASTG II)}}
\author[G. G. Nyambuya ]{G. G. Nyambuya\thanks{E-mail: gadzirai@gmail.com }, \\
}

\begin{document}

%\date{Received: / Accepted: }

%\pagerange{\pageref{firstpage}--\pageref{lastpage}} \pubyear{2009}

\maketitle
%\label{firstpage}

\begin{abstract}
{This reading expounds with expediency on the recently proposed Azimuthally Symmetric Theory \textit{of} Gravitation (ASTG) set-up earlier. At its inspection, it was demonstrated that the ASTG is  capable (among others  solar anomalies) of explaining the precession of the perihelion of solar planets.  In the present, we show that the ASTG is capable of explaining the puzzling observations of flyby anomalies, \textit{i.e.} the anomalous asymptotic speed increases of the osculating hyperbolic speed excess. It is shown that these flyby anomalies occur naturally in the ASTG. We derive the empirical formula proposed by  Anderson \textit{et al.} in $2008$, which up to now has no physical or foundational basis except that experience suggest it. If the ASTG model is correct, then for the first time the Anderson \textit{et al.} formula is given a physical meaning.} 
\end{abstract}


\section{Introduction}

Earth flyby anomalies have become a puzzling phenomenon. An Earth flyby anomaly is not just an unexpected  increase in the outgoing osculating hyperbolic excess speed but also an asymptotic speed increase at the perigee during Earth flybys of spacecraft. In general a flyby anomaly is an unexpected  increase in the outgoing osculating hyperbolic excess speed and as-well an asymptotic speed increase at the perigee during a flyby of a spacecraft past a planet for the purposes of gravity assist maneuver. This anomaly has been observed for spacecrafts sent to probe the secrets of deep space as they fly past the Earth as a shift in the   ranging  and Doppler data. For these spacecrafts, along their hyperbolic trajectory on their incoming path as they approach the Earth with a speed $v_{i}$ and when they exit at a speed of $v_{o}$, spherically symmetric Newtonian and Einsteinian gravitation dictates that $v_{i}= v_{o}$. In violation of this, observations give a completely different and surprising result that has baffled European Space Agency (ESA) and the National Aeronautic Space Administration (NASA) scientists for quite sometime now, \textit{i.e.} they [observations] reveal that $v_{i}\neq v_{o}$. Hence, the incoming kinetic energy of the spacecraft is less/greater than the outgoing kinetic energy of that spacecraft. 

Also, as the spacecraft reach their perigee -- that is, their distance of closest approach to planet Earth, it has been observed that these spacecrafts experience a hitherto unknown, mysterious and unexplained asymptotic speed increase. All this has come from the telemetry received from the spacecrafts. When the shift in the Doppler and the ranging data is interpreted, flyby anomalies are a very small -- \textit{albeit}, very significant unaccounted speed increase of up to $13.46\, \textrm{mm}/\textrm{s}$ at perigee. The first flyby anomaly was noticed during a very careful inspection of Doppler data shortly after the Earth flyby of the Galileo spacecraft on $8$ December $1990$ AD. While the Doppler residuals (observed minus computed data) were expected to remain flat, the analysis revealed an unexpected $66\, \textrm{mHz}$ shift, which corresponds to a speed increase of $3.92\, \textrm{mm}/\textrm{s}$ at perigee. An investigation of this effect at the Jet Propulsion Laboratory (JPL), the Goddard Space Flight Center (GSFC) and the University \textit{of} Texas has not yielded a satisfactory explanation. It should be noted that no anomaly was detected after the second Earth flyby of the Galileo spacecraft in December $1992$ AD, because any possible velocity increase is believed to have been masked by atmospheric drag of the lower altitude of $303\, \textrm{km}$.

On $23$ January $1998$ AD, the Near Earth Asteroid Rendezvous (NEAR) spacecraft experienced an anomalous speed increase of $13.46\, \textrm{mm s}^{-1}$ after its Earth encounter. Cassini-Huygens gained about $0.11\, \textrm{mm s}^{-1}$ in August $1999$ AD and Rosetta $1.82\, \textrm{mm s}^{-1}$ after its Earth flyby in March $2005$ AD. An analysis of the MESSENGER spacecraft (studying Mercury) did not reveal any significant unexpected velocity increase. The last Earth flyby was that by Rosetta in $2009$ AD. As she (Rosetta) bed farewell to humanity on her third and final Earth encounter  at $08:45$ in the European morning of the  $13^{th}$ of November $2009$ AD, on her trajectory to rendezvous with Comet $67$P/Churyumov-Gerasimenko on $22$ May $2014$ AD, the ESA spacecraft approached the Earth  before entering the depths of space in which event she left her highly expectant ``onlookers'' disappointed. While her ``onlookers'' watched her in the operation center, she  approached and passed closest to Earth over the south of the island of Java, in Indonesia, at a speed of $13.34\, \textrm{km}/\textrm{s}$ relative to the Earth, and at a height of $2481\,\textrm{km}$ above its surface. In the operation center -- that is, in the European Space Operation Center from ESA in Darmstadt (Germany), nothing special happened at that key moment. No applauses, nor hugs from the  pregnant engineers, \textit{i.e.}  pregnant with expectations because everything had been planned to the minute and the millimeter weeks in advance and Rosetta did not yield any significant flyby anomaly as highly expected! This only helped to propel the puzzle to newer heights. What is their cause? Is anything wrong with our understanding of gravitation? What is going on? These are just a few of the plethora of questions that come to flood the seeking mind.

For example, prior to the much awaited Rosetta III flyby, researchers  \cite{anderson08}  deduced an empirical relationship from which they predicted a flyby anomaly of up to  about $1\,\textrm{mm}/\textrm{s}$ for the $13$ November $2009$ AD Rosetta Earth encounter. This did not happen. What was measured is something to the tune of $0.004\pm0.044\,\textrm{mm}/\textrm{s}$ which for all practical purposes is a null result. The empirical relationship that \cite{anderson08} found is:

\begin{equation}
\frac{\Delta v}{v}=\kappa_{A}\left(\cos\delta_{i}-\cos\delta_{o}\right), \label{anderson}
\end{equation}

where $\kappa_{A}=2\mathcal{R}_{\oplus}\omega_{\tiny \earth}/c=3.10\times10^{-6}$ is the \cite{anderson08} constant and $\omega_{\tiny \earth}=7.29\times10^{-5}\,\textrm{rad}/\textrm{s}$ (see \textit{e.g.} Stacey  $1992$, in \citealt{anderson08}) is the angular frequency of the Earth, $\mathcal{R}_{\tiny \earth}=6.40\times10^{6}\,\textrm{m}$  is the radius of the Earth, and $\delta_{i}$ and $\delta_{o}$ are the incoming and outgoing osculating asymptotic velocity vectors. The Anderson formula (\ref{anderson})  has up to now no substantial physical basics in that an acceptable/accepted physical theory is yet to furnish its very foundations.

The \cite{anderson08} relationship came about after realizing that the MESSENGER spacecraft had both approached and departed the Earth symmetrically about the equator (\textit{i.e.} it approached at latitude $31$ degrees north and;  departed at latitude $32$ degrees south). This was taken as a strong suggestion that the anomaly might be related to the Earth's rotation and this incoming and outgoing velocity vectors. As already said above, this led \cite{anderson08} to successfully seek an empirical relationship involving the incoming and outgoing declination angles of the orbit of the spacecrafts. 

This empirical relationship of \cite{anderson08}, as already said, suffers from the setback that it has no physical explanation. This reading seeks (and hopes) not only to give the \cite{anderson08} empirical relationship a foundational basics but to give a physical explanation of these seeming puzzling observations. It shall be demonstrated that flyby anomalies emerge naturally in the Azimuthally Symmetric Theory \textit{of} Gravitation (ASTG) (\citealt{nyambuya10a}). 

It is known not whether this phenomenon of flyby anomalies may be related to the Pioneer Anomaly. Bonafide -- there is a significant number of researchers who (strongly) feel and suspect that these two phenomenon may very well be related. In its bare form, \textit{i.e.} original form, the ASTG is unable to account for the Pioneer Anomaly. It can be demonstrated that the ASTG model can in principle explain the Pioneer Anomaly if one adjusts the initial conditions of the ASTG model. We are not going to present this proposal of the extended ASTG model because we are currently at work on it and once we are certain of its correctness, we will forward this idea for publication. To evaluate this idea, that is, gain confidence that this result may be correct, one will need the complete/partial set of the Pioneer ephemerids. With this, one will be able to know whether the this extended ASTG model that we have in mind can face up with experience.

The synopsis of this reading is as follows. In the subsequent section, we present evidence pointing to the fact that the choice of the $\lambda$'s that we made in \cite{nyambuya10a} is good as it appears to be strongly backed by physical evidence. In \S (3), we present the main findings of the present reading, namely that the ASTG is able to explain reasonably well the puzzling flyby anomalies and in the penultimate, \textit{i.e.} \S (4), we give a general discussion and make our conclusions.

\section[\textbf{The Undetermined Constants $\lambda_{\ell}$}]{The Undetermined Constants $\lambda_{\ell}$}

As already stated in \cite{nyambuya10a}, one of the draw backs of the ASTG is that it is heavily dependent on observations \textit{for} the values of $\lambda_{\ell}$ have to be determined from observations. Without knowledge of the $\lambda_{\ell}'s$, one is unable to produce the hard numbers required to make any numerical quantifications. Clearly, a theory incapable of making any numerical quantifications is -- in the physical realm, useless. To avert this, already in \cite{nyambuya10a}, the determined solar values of the $\lambda_{\ell}'s$ have been used  to make what appears to be a \textit{reasonable suggestion}. It has been suggested that:

\begin{equation}
\begin{array}{l}
\lambda_{\ell}=\left(\frac{(-1)^{\ell+1}}{\left(\ell^{\ell}\right)!\left(\ell^{\ell}\right)}\right)\lambda_{1}\,\,\,\,\,\,\,\,\,\,\,\,\,\,\,\,\,\,\,\,\,\,\,\,\,\,\,\,\,\,\,\,\,\,\,\,\,\,\,\,\,\,\,\,\,\,\,\,\,\,\,\,\,\,\,\,\,\,\,\,\,\,\,\,\,\,\,\,\,\,\,\,\,\,\,\,\,\,\,\,\,\dots\textbf{(a)}\\
\\
\lambda_{\ell} =-\left(\frac{-1}{\left(\ell^{\ell}\right)!}\right)^{\ell}(\lambda_{1})^{\ell} \,\,\,\,\,\,\,\,\,\,\,\,\,\,\,\,\,\,\,\,\,\,\,\,\,\,\,\,\,\,\,\,\,\,\,\,\,\,\,\,\,\,\,\,\,\,\,\,\,\,\,\,\,\,\,\,\,\,\,\,\,\,\,\,\,\,\,\,\,\,\,\,\dots\,\,\textbf{(b)}
\end{array}
\label{oldlambda}
\end{equation}

\noindent This suggestion meets the intuitive requirements stated in \cite{nyambuya10a}. The second suggestion is new -- that is (\ref{oldlambda}). We happen to find that this same form of the $\lambda$; to second order, does meet the same requirements as the initial proposal. We shall however stick to the initial proposal made in \cite{nyambuya10a} and when the situation arises where we may need the second form, we will take it up. If $\lambda_{\ell}$ is given by the first form, then we should be able to obtain a more accurate value of $\lambda_{1}^{\odot}$. To do this we go back to equation (47) in \cite{nyambuya10a}, that is:

\begin{equation}
\mathscr{P}_{p}=\mathscr{A}_{p}\lambda^{\tiny \odot}_{1}+\mathscr{B}_{p}\lambda^{\tiny \odot}_{2},
\end{equation}

where the symbols are defined therein (\citealt{nyambuya10a}). From (\ref{oldlambda}), it follows that $\lambda^{\tiny \odot}_{2}=\lambda^{\tiny \odot}_{1}/96$ and substituting this into the above, one is led to:

\begin{equation}
\mathscr{P}_{p}=\lambda^{\tiny \odot}_{1}\left(\mathscr{A}_{p}-\mathscr{B}_{p}/96\right).
\end{equation}

Setting $X_{p}=\left(\mathscr{A}_{p}-\mathscr{B}_{p}/96\right)$,  implies $\mathscr{P}_{p}=\lambda^{\tiny \odot}_{1}X_{p}$ and since $\mathscr{P}_{p}$ and $X_{p}$ are known and $\lambda^{\tiny \odot}_{1}$ is unknown, a plot of $\mathscr{P}_{p}$ vs $X_{p}$ should produce a straight line whose slope is $\lambda^{\tiny \odot}_{1}$. The values $\mathscr{A}_{p}, \mathscr{B}_{p}, \mathscr{P}_{p}$ and $X_{p}$ are tabled in table \ref{graph:lambda} and the corresponding graph is plotted in figure \ref{perishift-graph}.  From the graph, we get:

\begin{equation}
\lambda^{\odot}_{1}=21.00\pm4.00.
\end{equation}

Obviously, this value $\lambda^{\odot}_{1}=21.00\pm4.00$ is more accurate than our earlier value  $\lambda^{\odot}_{1}=24.00\pm7.00$ hence we now adopt the former. The fact that we are able to obtain a very good linear graph as shown in figure \ref{perishift-graph}, this points to the fact that our choice of the $\lambda$'s is good. This exercise, to demonstrate that our choice of the parameter $\lambda$ is a good one, is the main thrust of the present section. We believe that demonstrating this fact using a graph gives impetus to our choice.

Now, we move onto make a further suggestion on the parameter $\lambda$. This is all in the effort of moving closer to resolving the ``ASTG Constants Problem''.  This suggestion shall be taken up more seriously in the reading \cite{nyambuya10b}. If these $\lambda$'s are to be given by (\ref{oldlambda} a), then, there is just one unknown parameter and this parameter is $\lambda_{1}$. The question is what does this depend on? We strongly feel/believe that $\lambda_{1}$ is dependent on the spin angular frequency and the radius of the gravitating body in question and our reasons are as follows.

The ASTG has been shown in \cite{nyambuya10b} to be able to explain outflows as a gravitational phenomenon. Pertaining to their association with star formation activity, it is believed that molecular outflows are a necessary part of the star formation process because their existence may explain the apparent angular momentum imbalance. It is well known that the amount of initial angular momentum in a typical star-forming cloud core is several orders of magnitude too large to account for the observed angular momentum found in formed or forming stars (see \textit{e.g.} \citealt{larson03}). The sacrosanct Law \textit{of} Conservation of angular momentum informs us that this angular momentum can not just disappear into the oblivion of interstellar spacetime. So, the question is where does this angular momentum go to? It is here that outflows are thought to come to the rescue as they can act as a possible agent that carries away the excess angular momentum. Whether or not this assertion is true or may have a bearing with reality, no one knows as verifying this is a mammoth task.

This angular momentum, if it where to remain as part of the nascent star, it would, \textit{via} the  strong centrifugal forces (the centrifugal acceleration is given by: $a_{c}=\omega^{2}_{star}\mathcal{R}_{star}$.),  tear the star apart.  This however does not explain, why they [outflows]  exist and how they come to exist but simple posits them as a vehicle needed to explain the mystery of ``{The Missing Angular Momentum Problem}'' in star forming systems and the existence of stars in their intact and compact form as firery balls of gas.

We draw  from the tacit thesis ``that outflows possibly save the star from the detrimental centrifugal forces'',  the suggestion that $\lambda_{1}\propto (a_{c})^{\zeta_{0}}$ where $\zeta_{0}$ is a pure constant that must be universal, that is, the same for all gravitating systems. This suggestion leads us to:

\begin{equation}
\lambda_{\ell}=\left(\frac{(-1)^{\ell+1}}{\left(\ell^{\ell}\right)!\left(\ell^{\ell}\right)}\right)\left(\frac{a_{c}}{a_{*}}\right)^{\zeta_{0}}\label{newlambda1},
\end{equation}

Knowing the solar values of $\lambda_{1}$ and as-well the value of $\zeta_{0}$, one is lead to: $a_{*}=\omega^{2}_{\odot}\mathcal{R}_{\odot}(\lambda_{1}^{\odot})^{-\frac{1}{\zeta_{0}}}$. As will be demonstrated in \cite{nyambuya10b}, the term $\lambda_{1}$ controls outflows. Given that $\lambda_{1}$ controls outflows and that outflows possibly aid the star in shedding off excess spin angular momentum, the best choice\footnote{We speak of ``choice'' here as though the decision is ours on what this parameter must be. No, the decision was long made by \textit{Nature}, ours is to find out what choice \textit{Nature} has made. That said,  we should say that, this ``choice'' is made with expediency -- \textit{i.e.}, this choice which is based on intuition, is to be measured against experience.} for this parameter is one that leads to these outflows responding to the spin of the star and as well the centrifugal forces generated by this spin in such a way that the star is able to shed off this excess spin angular momentum. So, what leads us to this proposal $\lambda_{1}\propto (a_{c})^{\zeta_{0}}$ is the aforesaid. This will became clear in \cite{nyambuya10b}. In brief we simple have this to say; since the spin generates centrifugal forces which would tear a star, and knowing that if $\lambda_{1}\propto (a_{c})^{\zeta_{0}}$, it is possible that when the spin reaches a critical state (determined by $a_{*}$) when these centrifugal forces are able to tear the star apart, the star switches on its polar repulsive gravitation field so as to get reed of this excess spin angular momentum.

\begin{table}

\centering
\caption[Graph to Deduce $\lambda_{1}^{\odot}$]{Column $(1)$ gives the planet while columns $(3\,to\,5)$ give values of $\mathscr{A}_{p}, \mathscr{B}_{p}, \mathscr{P}_{p}$ and $X_{p}$ for the corresponding planets respectively.}
\begin{tabular}{l l l l l }
\hline
\textbf{Planet}  & $\mathscr{A}$ & $\mathscr{B}$ & $\mathscr{P}$  & $X$ \\
\hline\hline
\textbf{Mercury} & $3.50\times10^{0}$   & $1.72\times10^{2}$	 &  $43.1000\pm 0.5000$  & $1.71\times10^{0}$\\
\textbf{Venus}   & $5.19\times10^{-1}$  & $2.88\times10^{1}$   & 	$\,\,\,8.0000\pm 5.0000$   & $4.89\times10^{-1}$\\
\textbf{Earth}   & $1.57\times10^{-1}$  & $3.80\times10^{-1}$  &	$\,\,\,5.0000\pm 1.0000$   & $1.53\times10^{-1}$\\
\textbf{Mars}    & $7.02\times10^{-2}$  & $2.43\times10^{-2}$  &	$\,\,\,1.3624\pm 0.0005$   & $7.00\times10^{-2}$\\
\textbf{Jupiter} & $3.02\times10^{-3}$  & $1.00\times10^{-5}$  &	$\,\,\,0.0700\pm 0.0040$   & $3.32\times10^{-3}$\\
\textbf{Saturn}  & $7.59\times10^{-4}$  & $1.72\times10^{-7}$  &	$\,\,\,0.0140\pm0.0020$    & $7.93\times10^{-4}$\\
\hline
\end{tabular}

\label{graph:lambda}
\end{table}

Now, as will be seen in (\ref{earthlambda1}) and (\ref{earthlambda2}), depending on one's interpretation of our derived flyby equation (\ref{nanderson}), $\lambda_{1}^{\oplus}$ takes the value $15000\pm7000$ or $2000\pm800$ respectively. The value $\lambda_{1}^{\oplus}$ is the $\lambda_{1}$-value of the Earth. If $\lambda_{1}^{\oplus}=15000\pm7000$, then:

\begin{equation}
\frac{\lambda_{1}^{\oplus}}{\lambda_{1}^{\odot}}=\frac{15000\pm7000}{21.00\pm4.00}=800\pm500,\label{ratio1}
\end{equation}

and this would imply that $\zeta_{0}=5.0$, since: 

\begin{figure*}
\centering
\includegraphics[scale=0.42]{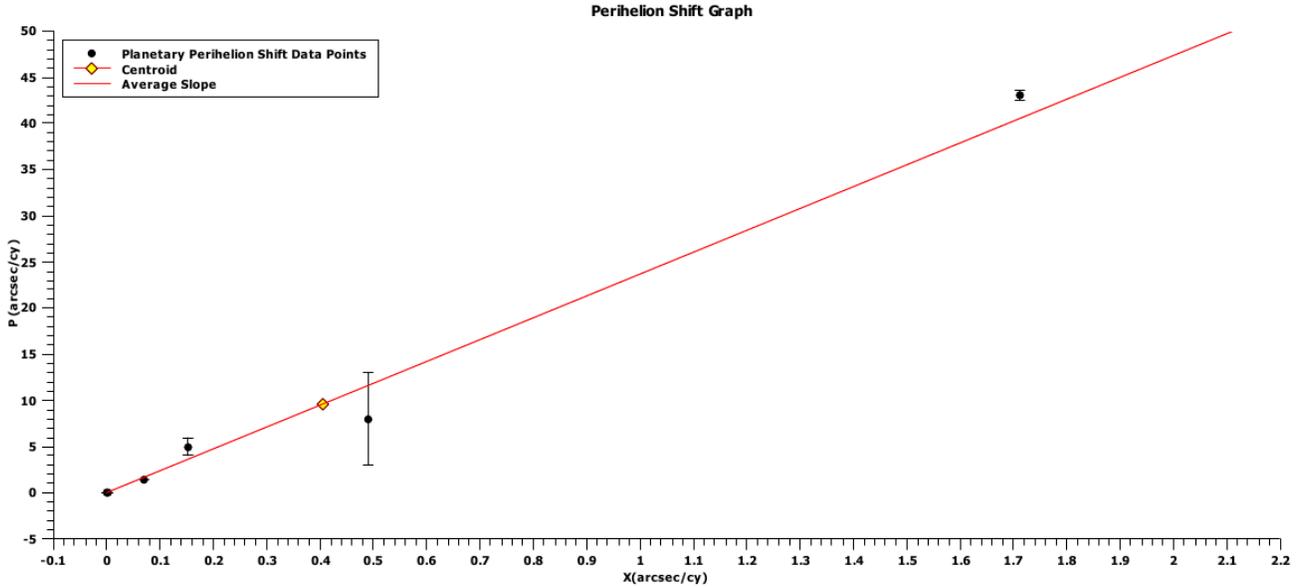}
\caption[\textbf{Perihelion Shift Graph}]{A plot of the perihelion shift data points of solar planetary orbits. This graph is (in our opinion and view) the most convincing piece of evidence yet, that the choice of the $\lambda$'s made in \cite{nyambuya10a}, may very well be a correct one.}
\label{perishift-graph}
\end{figure*}

\begin{equation}
\left(\frac{\omega_{\oplus}^{2}\mathcal{R}_{\oplus}}{\omega_{\odot}^{2}\mathcal{R}_{\odot}}\right)^{5.0}\simeq800,
\end{equation}

where $\omega_{\oplus}$ is the angular frequency and $\mathcal{R}_{\oplus}$ the radius of the Earth respectively. If $\lambda_{1}^{\oplus}=2000\pm800$, then:

\begin{equation}
\frac{\lambda_{1}^{\oplus}}{\lambda_{1}^{\odot}}=\frac{2000\pm800}{21.00\pm4.00}=100\pm60,\label{ratio2}
\end{equation}

and this would imply that $\zeta_{0}\sim2.5$, since:

\begin{equation}
\left(\frac{\omega_{\oplus}^{2}\mathcal{R}_{\oplus}}{\omega_{\odot}^{2}\mathcal{R}_{\odot}}\right)^{2.5}\simeq118.
\end{equation} 

From all this, one can deduce that:

\begin{equation}
\zeta_{0}=3.75\pm1.25.\label{zeta}
\end{equation}

If  $\lambda_{1}$ where to take the least value, \textit{i.e.} $\lambda_{1}=2000\pm800$, then (as one can verify for themselves), a $1\%$ change in the period leads to a $1\%$ decrease in the spin angular frequency and in turn a $1\%$ change in the spin angular frequency leads to to a $7.3\%$ change in the value of $\lambda_{1}$ and a $1\%$ change in the radius of the gravitating body in question, leads to a $3.5\%$ change in the value of $\lambda_{1}$. The point we want to bring home is that, if $\zeta_{0}$ is in our suspected range of $3.75\pm1.25$, $\lambda_{1}$ is sensitive to the changes in the spin angular frequency and as well changes in the size of the gravitating body. This would mean for example that slit variations in the period would lead to variation in $\lambda_{1}$ and in the case of the Sun whose radius varies periodically, $\lambda_{1}$ must vary periodically in responce to this. 

\noindent  We are of the view that our thoughts as presented herein on what the $\lambda_{1}$-parameter  ought to be, must at the very least, give one hope that this problem of the unknown $\lambda$-parameters is within reach. We should say, that the way that we are going round this problem of the $\lambda_{1}$-parameter is not rigorous but is largely dependent on intuition, which for some reason, we happen to trust. We fully understand the fact that intuition can be very wrong and misleading, but here we are developing something utterly new -- we are trading in waters never chattered before; we are moving in the dark; hence, we must use our intuition to the best of our abilities. We will seek evidence to try and backup our assertions in \cite{nyambuya10b}. It is very important to state that our musings on what the $\lambda_{1}$-parameter ought to be, does not affect at all the findings of this reading namely that the ASTG is able to explain the flyby anomalies. Actually, this reading would do without the present section. We have presented this section only as an effort to make strides in resolving the ``The ASTG Constants Problem''.

\section[\textbf{Anomalous Speed Changes of Spacecraft at Infinity}]{Anomalous Speed Changes of Spacecraft at Infinity}

\begin{figure}
\centering
\includegraphics[scale=0.26]{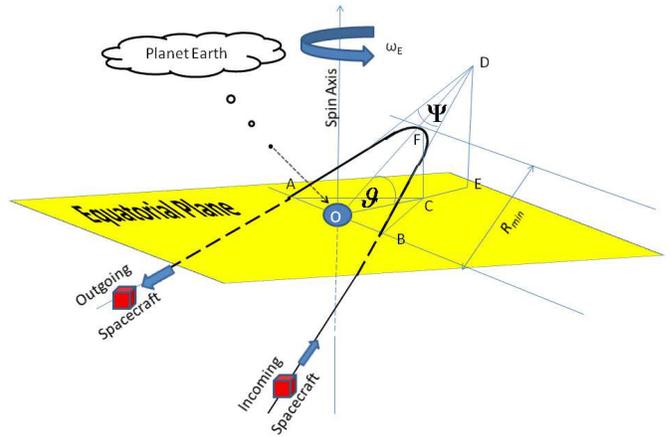}
\caption[\textbf{Schematic Diagram of a Spacecraft Flyby}]{Schematic diagram showing the geometry of the orbit of a spacecraft making a planetary flyby.}
\label{flybydiag}
\end{figure} 

\begin{figure}
\centering
\includegraphics[scale=0.25]{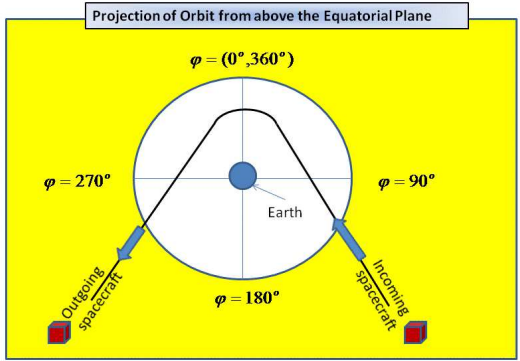}
\caption[\textbf{Digram Showing the Equatorial View of  Spacecraft Flyby Orbits}]{An illustration showing the equatorial view of  spacecraft flyby orbits.}
\label{eviewdiag}
\end{figure}

In \cite{nyambuya10a} (equation 70), it has been shown that the orbit equation that emerges from the ASTG model is:

\begin{equation}
\frac{l}{r}=1+\epsilon_{N}e^{k\varphi}\cos([\eta_{2}+\eta_{3}]\varphi).
\end{equation} 

For an object such as the Earth $\eta_{2}+\eta_{3}\sim1$, hence the above equation for Earth orbits is given by:

\begin{equation}
\frac{l}{r}=1+\epsilon_{N}^{\oplus}e^{k_{\oplus}\varphi}\cos\varphi\label{orb1-fba},
\end{equation}

where $\epsilon_{N}^{\oplus}$ is the Newtonian eccentricity for Earth orbits and this is given by:

\begin{equation}
\epsilon_{N}^{\oplus}=\left(\frac{v^{2}_{\infty}}{G\mathcal{M}_{\oplus}/\mathcal{R}_{min}}\right),
\end{equation}

where $\mathcal{R}_{min}$ is the distance of closest approach and:

\begin{equation}
k_{\oplus}=\frac{\lambda_{1}^{\oplus}}{2}\left(\frac{G\mathcal{M}_{\oplus}}{c^{2}\mathcal{R}_{min}}\right)\sin\theta.
\end{equation}

For an explanation of the symbols of all the equations above, we direct the reader to \cite{nyambuya10a}. Actually, to make sense of the present reading, the reader will have to first go through \cite{nyambuya10a}. Having gone through \cite{nyambuya10a}, the next thing is to understand the geometry of the orbit itself.

Since $\epsilon_{N}^{}\propto v^{2}_{\infty}$ and given that $v_{\infty}^{i}\neq v_{\infty}^{o}$ where $(v_{\infty}^{i}, v_{\infty}^{o})$ are the incoming and outgoing osculating hyperbolic excess speed respectively, the points to the fact that $\epsilon_{N}^{i}\neq\epsilon_{N}^{o}$. But how does this come about that $\epsilon_{N}^{i}\neq\epsilon_{N}^{o}$? To answer this question we have to look into the orbits and the equation of the orbit.

First, let us go to figure \ref{eviewdiag}. For an unbound orbit $\varphi:\,\,(0\degree\leq \varphi\leq 360\degree)$ and at the perigee $\varphi=(0\degree,360\degree)$. So, when the spacecraft reaches the perigee, it encounters two different values for $\varphi$, \textit{i.e.}: $\varphi=(0\degree,360\degree)$. The functions ($\sin\varphi,\cos\varphi$) do not have a problem with this apparent asymptotic change in the $\varphi$-value, that is from the value $\varphi=0\degree$ to $ \varphi = 360\degree$ (or  $360\degree \longmapsto0\degree$, this depends on the direction from which the spacecraft approaches the perigee) these function ($\sin\varphi,\cos\varphi$) are smooth continuous. For the pre-perigee orbit, we have $(0\degree\leq\varphi\leq180\degree: \varphi\,\,\textrm{moves}\,\,\textrm{from}\,\, 180\degree\longrightarrow0\degree)$ and for the post-perigee orbit, we have $(360\degree\leq\varphi\leq180\degree: \varphi\,\,\textrm{moves}\,\,\textrm{from}\,\, 360\degree\longrightarrow180\degree)$.

At the perigee, a function like $e^{k\varphi}$ will have a problem  since there at the perigee there exists two values of $\varphi=(0\degree,360\degree)$. It would have to jump from $1\longmapsto e^{2\pi k}$. It is here that we expect the speed jumps at the asymptote to have their origins. We shall not look into the speed jumps at the perigee. Clearly, the fact that for the pre-perigee orbit, we have $(0\degree\leq\varphi\leq180\degree: \varphi\,\,\textrm{moves}\,\,\textrm{from}\,\, 180\degree\longrightarrow0\degree)$ and for the post-perigee orbit, we have $(360\degree\leq\varphi\leq180\degree: \varphi\,\,\textrm{moves}\,\,\textrm{from}\,\, 360\degree\longrightarrow180\degree)$ means the function $e^{k\varphi}$ is not symmetric about the perigee. This means the orbit itself is not symmetric about the perigee as is the case in spherically symmetric Newtonian gravitation. This asymmetry is the origins of the  outgoing osculating hyperbolic excess speed. In \cite{nyambuya10a} where the ASTG was first laid down, we did show there-in that the eccentricity of a orbit has an additional term $e^{k\varphi}$ such that $\epsilon=\epsilon_{N}e^{k\varphi}$ where for the Earth, hence this asymmetric will lead to the eccentricity of the incoming and outgoing orbit to be different, hence the  outgoing osculating hyperbolic excess speed. We have justified our assertion that $\epsilon_{N}^{i}\neq\epsilon_{N}^{o}$. 

For bound orbits such as the Earth in its orbit around the Sun, $(-\infty\leq\varphi\leq+\infty)$, the meaning of which is that $\varphi$ is continuous at the perigee. Thus, this strange behavior seen in flybys is, in accordance with the ASTG, not expected to occur.

Now, we move on the main task -- that of showing that the ASTG does explain the speed increase in the outgoing osculating hyperbolic excess speed. For the geometry of the orbit, we have made the illustration in figure \ref{flybydiag}. At the perigee, we must have $\varphi=0$, and for this to be so, we must have $\varphi=\alpha-\alpha_{prg}$ where $\alpha$ is the RA of the spacecraft at any given point on the orbit and $\alpha_{prg}$ is RA angle at the perigee. At the perigee, $\alpha=\alpha_{prg}$ hence $\varphi=\alpha_{prg}-\alpha_{prg}=0$. The polar coordinate system that we use here is the same as that defined in \cite{nyambuya10a}. Now for $\theta$, it is not difficult to see that $\theta=90\degree+\delta$ where $\delta$ is the DEC angle of the spacecraft at any given point on the orbit. Hence $(\theta,\varphi)=(90\degree+\delta, \alpha-\alpha_{prg})$.

\begin{table*}

\centering
\caption[Flyby Anomalies Parameters and Deduction of $a_{*}$]{Earth flyby parameters at the asymptotes of their orbits for Galileo, NEAR, Cassini, Rosetta, and MESSENGER spacecraft. Columns $(1), (2)\,\textit{\&}\,(3)$ gives the name of the spacecraft, the date it made its gravity assist maneuver and the Agency responsible for this spacecraft respectively. Columns $(4)\,\,\textrm{to}\,\,(7)$ gives the incoming and outgoing Right Ascension, the incoming and outgoing Declination angle respectively. Columns $(8)\,\,\textrm{to}\,\,(10)$ are the osculating hyperbolic excess velocity,   the altitude is referenced to an Earth geoid plus the radius of the Earth, and the change in osculating hyperbolic excess velocity. Column ($11$) is the $k_{A}$ value from the spacecraft data and the ASTG while data column ($12$) is the direct value of $\lambda_{1}^{\earth}$ calculated from equation (\ref{nanderson}).  The values of $\lambda_{1}^{\earth}$ in column ($11$) have been calculated from equation (\ref{nanderson}) by making  $\lambda_{1}^{\earth}$ the subject of the formula. The data in this table except for that in column $(11)$ \textit{\&} $(12)$, is adapted from \cite{anderson08}.}

\begin{tabular}{l c r c r r r r r r  r r}
\hline
\textbf{Spacecraft}  & \textbf{Date} & \textbf{Agency} &   $\alpha_{i}$ & $\alpha_{o}$ & $\delta_{i}$ & $\delta_{o}$ & $v_{\infty}$ &  $\mathcal{R}_{min}$ & $\Delta v^{obs}_{\infty}$ & $k_{A}$  & $\lambda_{1}^{\oplus}$ \\
                     &               &            &  ($1\degree$) & ($1\degree$) & ($1\degree$)  & ($1\degree$) &  ($\textrm{km}/\textrm{s}$)  & ($\textrm{km}$)  &  ($\textrm{mm}/\textrm{s}$) &  $(10^{-7})$\\
\hline\hline
\textbf{Galileo I}       &  $08/12/1990$ & NASA &  $266.76$ & $219.97$ & $12.52$   & $34.15$  & $8.949$  &  $7356$   &  $3.92\pm0.08$  &  $8.00\pm3.00$ & $2750\pm60$ \\
\textbf{Galileo II}      & $12/12/1992$ & NASA  &  $219.35$ & $174.35$ & $-34.26$ & $-4.87$  &  $8.877$ &  $6703$   &  $-4.60\pm1.00$    & $9.00\pm4.00$ & $2700\pm600$\\
\textbf{NEAR}            & $23/01/1998$ & NASA  &  $261.17$ & $183.49$ & $-20.76$  & $-71.96$ &  $6.851$ &  $6939$   &  $13.46\pm0.13$  &  $14.20\pm0.70$ & $1930\pm30$\\
\textbf{Cassini}         & $18/08/1999$ & NASA  &  $334.31$  & $352.54$ & $-12.92$  & $-4.99$  &  $1.601$ &  $7571$   &  $-2.00\pm1.00$  &  $3.00\pm2.00$ & $40000\pm20000$\\
\textbf{Rosetta I}       &  $04/03/2005$ & ESA  &  $346.12$ & $246.51$ & $-2.81$   & $-34.29$ &  $3.863$ &  $8354$   &   $1.80\pm0.05$  &  $15.10\pm0.70$ & $1750\pm50$\\
\textbf{M''NGER}       &  $02/08/2005$ & Private &  $292.61$ & $227.17$ & $31.44$   & $-31.92$ &  $4.056$ &  $8736$   &   $0.02\pm0.01$  &  $10.00\pm4.00$ & $900\pm400$\\
%\textbf{Rosseta II}      & $13/11/2007$ & ESA    & --  & -- & -- & -- & --                    & $3.863$  &  $11722$ &  $\sim0$         & --             \\
%\textbf{Rosseta III}     & $13/11/2009$ & ESA    & --  & -- & -- & -- & --                    & $3.863$  &  $8883$  &  $\sim0$ &  --             \\
\hline
\hline
\textbf{Mean}         & & & & & & & & & & $10.00\pm5.00$ & $2000\pm200$ \\
\textbf{Std. Dev.}    & & & & & & & & & & $4.00\pm2.00$ & $800\pm300$ \\
\hline
\end{tabular}
\label{table:flybyparms}
\end{table*}

\begin{figure*}
\centering
\includegraphics[scale=0.41]{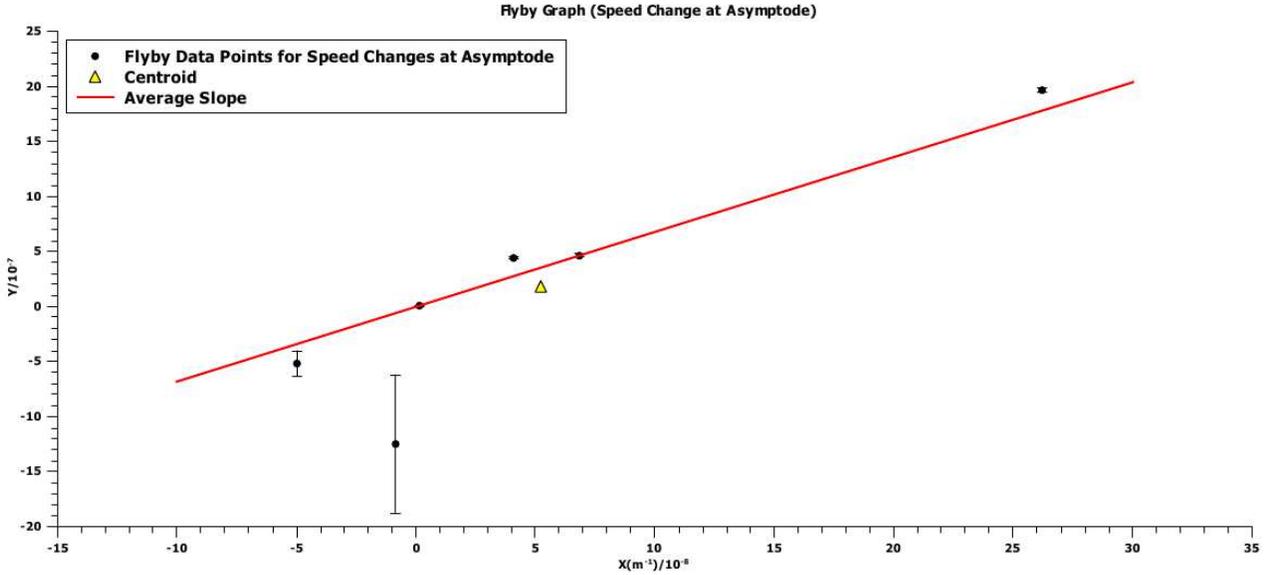}
\caption[\textbf{Asymptotic Speed Change Graph}]{A graph showing the asymptotic speed changes according to the ASTG.  Along the $x$-axis, the data points from left to right are for Cassini, Galileo II, Messenger, Galileo I,  Rosetta I and    and NEAR. The data point for Rosetta I. The Galileo II point falls the line of average slope. This point was not used in the computation for average value of the slope. The points used in the computation of the average slope are those for Galileo I and  Rosetta I. That of Galileo I was used for computing the maximum slope while that for  Rosetta I was used to compute the minimum slope.}
\label{fb-asymptodediag}
\end{figure*} 

Now, for the pre-perigee encounter, when $r=\infty$, $\varphi=\textrm{A}\hat{\textrm{C}}\textrm{B}/2=|\alpha_{i}-\alpha_{o}|/2$ and $\theta=90+\delta_{i}$ where the subscript ($i,o$) on the angles ($\delta,\alpha$) are labels to indicate that these angles are for to the incoming ($i$) and ($o$) outgoing RA and DEC angles. Substituting these parameters in (\ref{orb1-fba}), we are led to: 

\begin{equation}
0=1+\epsilon_{N}^{i}e^{k_{\oplus}^{i}|\alpha_{i}-\alpha_{o}|/2}\cos\left(\frac{|\alpha_{i}-\alpha_{o}|}{2}\right)\label{orb3-fba}.
\end{equation}

Likewise, for the post-perigee encounter, when $r=\infty$, $\varphi=\textrm{A}\hat{\textrm{C}}\textrm{B}/2=|\alpha_{i}-\alpha_{o}|/2$ and $\theta=90+\delta_{o}$, this means: 

\begin{equation}
0=1+\epsilon_{N}^{o}e^{k_{\oplus}^{o}|\alpha_{i}-\alpha_{o}|/2}\cos\left(\frac{|\alpha_{i}-\alpha_{o}|}{2}\right)\label{orb4-fba}.
\end{equation}

Now, subtracting (\ref{orb3-fba}) from (\ref{orb4-fba}) and then dividing the resulting equation by $\cos\left(|\alpha_{i}-\alpha_{o}|/2\right)$, one is led to: $\epsilon_{N}^{i}e^{k_{\oplus}^{i}|\alpha_{i}-\alpha_{o}|/2}-\epsilon_{N}^{o}e^{k_{\oplus}^{o}|\alpha_{i}-\alpha_{o}|/2}=0$. Since $k_{\oplus}|\alpha_{i}-\alpha_{o}|/2$ is small, the approximation $e^{k_{\oplus}|\alpha_{i}-\alpha_{o}|/2}\simeq1+k_{\oplus}|\alpha_{i}-\alpha_{o}|/2$ holds.  Using this approximation into the equation: $\epsilon_{N}^{i}e^{k_{\oplus}^{i}|\alpha_{i}-\alpha_{o}|/2}-\epsilon_{N}^{o}e^{k_{\oplus}^{o}|\alpha_{i}-\alpha_{o}|/2}=0$, one is led to: $(\epsilon_{N}^{i}-\epsilon_{N}^{o})/\epsilon_{N}^{i}=[k_{\oplus}^{i}-(\epsilon_{N}^{o}/\epsilon_{N}^{i})k_{\oplus}^{o}]|\alpha_{i}-\alpha_{o}|/2$. First,  the approximation $(\epsilon_{N}^{o}/\epsilon_{N}^{i})\sim1$ holds hence $(\epsilon_{N}^{i}-\epsilon_{N}^{o})/\epsilon_{N}^{i}=[k_{\oplus}^{i}-k_{\oplus}^{o}]|\alpha_{i}-\alpha_{o}|/2$. It is not difficult to deduce that: $(\epsilon_{N}^{i}-\epsilon_{N}^{o})/\epsilon_{N}^{i}=(v_{i,\infty}^{2}-v_{o,\infty}^{2})/v_{i,\infty}^{2}=\Delta \mathcal{K}/\mathcal{K}_{i}$ where $\Delta \mathcal{K}=\mathcal{K}_{i}-\mathcal{K}_{o}$ is the change in the kinetic energy of the spacecraft and $\mathcal{K}_{i}$ and $\mathcal{K}_{o}$ are the incoming and outgoing kinetic energies of the spacecraft at the asymptotes. Since $\Delta \mathcal{K}/\mathcal{K}_{i}=2\Delta v_{\infty}/v_{\infty}$, this means: $(\epsilon_{N}^{i}-\epsilon_{N}^{o})/\epsilon_{N}^{i}=2\Delta v_{\infty}/v_{\infty}$. For $k_{\oplus}^{i}-k_{\oplus}^{o}$ we have $k_{\oplus}^{i}-k_{\oplus}^{o}=(\lambda_{1}^{\oplus}/2)(G\mathcal{M}/c^{2}\mathcal{R}_{min})[\sin(90\degree+\delta_{i})-\sin(90\degree+\delta_{o})]$ hence $k_{\oplus}^{i}-k_{\oplus}^{o}=(\lambda_{1}^{\oplus}/2)(G\mathcal{M}/c^{2}\mathcal{R}_{min})[\cos\delta_{i}-\cos\delta_{o}]$.  Now effecting all this into: $(\epsilon_{N}^{i}-\epsilon_{N}^{o})/\epsilon_{N}^{i}=[k_{\oplus}^{i}-(\epsilon_{N}^{o}/\epsilon_{N}^{i})k_{\oplus}^{o}]|\alpha_{i}-\alpha_{o}|/2$, one is led to:

\begin{equation}
\left(\frac{\Delta  v_{\infty}}{v_{\infty}}\right)=\lambda_{1}^{\oplus}\left(\frac{\pi|\alpha_{i}-\alpha_{o}|G\mathcal{M}_{\oplus}}{1440c^{2}\mathcal{R}_{min}}\right)\left(\cos\delta_{i}-\cos\delta_{o}\right),\label{nanderson}
\end{equation}

which has the same form as the \cite{anderson08} formula (\ref{nanderson}).  Comparison of the above with (\ref{anderson}), gives:

\begin{equation}
k_{A}=\lambda_{1}^{\oplus}\left(\frac{\pi |\alpha_{i}-\alpha_{o}|G\mathcal{M}_{\oplus}}{1440c^{2}\mathcal{R}_{min}}\right).\label{nanderson-const}
\end{equation}

In the above and in (\ref{nanderson}), we have inserted the factor $\pi/180$ because the angles $\alpha$ are in degrees hence the factor $\pi/1440=\pi/(8\times180)$.

There is one unknown ($\lambda_{1}^{\oplus}$) in equation (\ref{nanderson}) thus we can calculated this given $\alpha_{i},\alpha_{o}, \delta_{i},\delta_{o}$ and $\mathcal{R}_{min}$. These values are given in table \ref{table:flybyparms}. If we set:

\begin{equation}
Y=\left(\frac{\Delta v_{\infty}}{v_{\infty}}\right)\,\,\, \textrm{and}\,\,\, X=\left(\frac{\pi\left(\sin\delta_{i}-\sin\delta_{o}\right)|\alpha_{i}-\alpha_{o}|}{180\mathcal{R}_{min}}\right),\label{xy1}
\end{equation}

then, a plot of $Y$ vs $X$ should yield an estimate of $\lambda_{1}^{\oplus}$ since the values of $G,\mathcal{M}_{\oplus}$ and $c$ are known. We have $Y=mX$ where the slope $m$ of this graph of $Y$ vs $X$ is:

\begin{equation}
m=\lambda_{1}^{\oplus}\left(\frac{G\mathcal{M}_{\oplus}}{8c^{2}}\right).
\end{equation}

We find from the graph in figure \ref{fb-asymptodediag} that $m=7.00\pm 4.00$. This slope value leads to: 

\begin{equation}
\lambda_{1}^{\oplus}=15000\pm 7000.\label{earthlambda1}
\end{equation}

This value assumes that $\lambda_{1}^{\oplus}$ is the same for all the flybys. As seen in figure \ref{fb-asymptodediag}, less the value for Galileo I, the rest of the values for other spacecrafts lie very close to the graph with the average slope. The graph of $Y$ vs $X$ is expected to pass through the point $(0,0)$. So, to obtained the average value of the slope, we computed from the data points on the $Y$ vs $X$ graph in figure \ref{fb-asymptodediag}  the maximum and the minimum slope and we took their average and for the error in this slope we computed the difference in these two values and divided by $2$ and from this we obtained the error in the slope. Judging from the graph in figure \ref{fb-asymptodediag},  we are of the view that this graph is acceptable linear relationship. This graph points to the ASTG as containing in it, a grain of truth to do with the flyby anomalies.

Given our thinking that $\lambda_{1}^{\oplus}$ should be dependent on the radius and as-well the period of the spin of the Earth and given that the spin of the Earth is not truly constant, then, we have a reason to believe that $\lambda_{1}^{\oplus}$ will not be the same for all the flybys as these flybys occur at different times when the Earth's spin is not the same. However, we know that this variation of the Earth day  is not so marked.  Given this, that the Earth day does not vary widely, it means we must not expect $\lambda_{1}^{\oplus}$ to vary widely as-well. In this case, (\ref{earthlambda1}) would be the most probable value of $\lambda_{1}^{\oplus}$.

If the Earth day did vary markedly, then $\lambda_{1}^{\oplus}$ would have to be calculated directly from (\ref{earthlambda1}). Presented in column ($12$) of table \ref{table:flybyparms} are the direct values of $\lambda_{1}^{\oplus}$ from the formula (\ref{earthlambda1}). These values have been obtained by making $\lambda_{1}^{\oplus}$ the subject of the formula and then inserting the relevant values from table \ref{table:flybyparms} in the resulting formula. The $\lambda_{1}^{\oplus}$-value of Galileo II is strangely high and we have excluded this from our calculations. We believe this high value clearly points to the fact that this interpretation of (\ref{nanderson}) to deduce $\lambda_{1}^{\oplus}$ is not correct as this would implies to marked variation in the Earth day.   With the Galileo II $\lambda_{1}^{\oplus}$-value excluded, one finds that:

\begin{equation}
\lambda_{1}^{\oplus}=2000\pm 800.\label{earthlambda2}
\end{equation}

The error in (\ref{earthlambda2}) is the standard deviation in the mean.

\section{Discussion and Conclusion}

The fact that we have been able to give a physical explanation behind the \cite{anderson08} formula from the well known and well accepted Poisson equation strongly suggests that the ASTG has in it some element of truth to do with flyby anomalies.  Clearly there is need for researchers to look into the ASTG as this theory flows from a natural solution of the well known  Poisson equation.  That we understand the Poisson equation is something almost taken for granted. Surely and clearly, we have made not any modification(s) to the Poisson equation but merely took its natural azimuthal solution and applied it to the scenario of a gravitational field of a spinning body.

The present attempt to explain flyby anomalies from conventional physics -- if successful, it would be the first such. \cite{lammerzahl06} have studied and dismissed by  a number of mundane causes for the Earth flyby anomalies, including Earth atmosphere, ocean tides, solid Earth tides spacecraft charging, magnetic moments, Earth albedo, solar wind, coupling of Earth's spin with rotation of the radio wave, Earth gravity, and relativistic effects predicted by Einstein's theory. All these potential sources of systematic error, and more, are modeled in the Orbit Determination Program (ODP). None of these phenomena seem able to account for these observed anomalies (\citealt{lammerzahl06}).

With most mundane causes having been ruled out  (\textit{e.g.} \citealt{lammerzahl06}), speculation becomes the order of the day. For example, \cite{adler09} tries to use darkmatter to solve this problem and \cite{mcculloch08} uses the idea that the inertia of matter is affected by a change in the acceleration. Other attempts invoke the gravitomagnetic field \textit{e.g.} \cite{iorio09} and other more realistic attempts are that there exists an energy transfer between the spacecraft and the planet \textit{e.g.} \cite{anderson06}. 

We should mention that when \cite{anderson08} proposed their empirical formula, they conjectured that flyby anomalies must be related to the spin of the Earth. This is in line with the ASTG, since the azimuthally symmetric gravitational field has everything to do with the spin angular momentum of the Earth. 

Before we close this reading, it is important that we mention that from the ASTG model, we have presented herein an explain of flyby anomalies for the change in the outgoing osculating hyperbolic excess speed but not for the asymptotic speed increase at the perigee. The reason for this is that we find that to explain the speed changes at the perigee, this will only be possible with the extended ASTG model which is currently under construction as mention in the penultimate of the in the introduction of this reading. In a future reading, we will present our findings on this.

In closing, allow us to say the following, that; the formula we obtained for predicting the anomalous increase in hyperbolic excess speed is  congruent to that of \cite{anderson08}. Additionally, prior to the present reading, \textit{i.e.} from \cite{anderson08}, only two parameters appeared to matter in as far as predicting the observed anomalous speed increase of the spacecraft at infinity and these are the incoming hyperbolic excess speed and the declination angle (incoming and outgoing). In the present, we have added  three more and these are the incoming and outgoing RA-angles ($\alpha_{i},\alpha_{o}$) and the perigee distance ($\mathcal{R}_{min}$, measured from the center of the Earth). As these parameters have been used to determine the flyby anomalies, it appears to us highly unlikely that they behave so well by chance; against this probability, we strongly believe we herein have a theory that strongly appears to contain in it, an element of truth.  Perhaps, researchers should excogitate on the possibility that the gravitational field of a spinning body is not Newtonian, but azimuthally symmetric as laid down in \cite{nyambuya10a}, \cite{nyambuya10b} and in the present.

%\section*{ Acknowledgments}

%I am grateful to my brother George and his wife Samantha for their kind hospitality they offered while working on this reading and to Isak D. Davids \& M. Christina Eddington for proof reading the grammar and spelling. Further, I am grateful to the unanimous reviewers, their invaluable criticism that has help in the refinement of the arguments presented. Last and certainly not least, I am very grateful to my Professor, D. Johan van der Walt and Professor Pienus Kobus, for the strength and courage that they have given me.

%\label{lastpage}
\end{document}